\documentclass[onecolumn]{aa}
\usepackage{amsmath}
\usepackage{amssymb}
\usepackage{graphics}
\usepackage{hyperref}
\usepackage[varg]{txfonts}
\usepackage{natbib}
\usepackage[utf8]{inputenc}

\usepackage{soul}

\begin{document}

\title{High Excitation Emission Line Nebula associated with an Ultra Luminous X-ray Source at $z=$ 0.027 in the \textit{AKARI} North Ecliptic Pole Deep Field.}

\author{J. Díaz Tello\inst{\ref{1},\ref{8}}\and T. Miyaji\inst{\ref{1},\ref{8}}\and T. Ishigaki\inst{\ref{2}}\and M. Krumpe\inst{\ref{3}}\and Y. Ueda\inst{\ref{4}}\and H. Brunner\inst{\ref{5}}\and T. Goto\inst{\ref{6}}\and H. Hanami\inst{\ref{2}}\and Y. Toba\inst{\ref{7}}.}

\institute{
Instituto de Astronomía sede Ensenada, Universidad Nacional Autónoma de México, Ensenada, Mexico \email{jadiazt@astrosen.unam.mx}\label{1}\and
Physics Section, Faculty of Science and Engineering, Iwate University, Morioka, Japan\label{2}\and
Leibniz Institut für Astrophysik Potsdam, An der Sternwarte 16, 14482 Potsdam, Germany\label{3}\and
Department of Astronomy, Kyoto University, Kyoto 606-8502, Japan\label{4}\and
Max-Planck-Institut für extraterrestrische Physik, Garching, Germany\label{5}\and
National Tsinghua University, Hsingchu, Taiwan\label{6}\and
Academia Sinica Institute of Astronomy and Astrophysics, PO Box 23-141, Taipei 10617, Taiwan\label{7}\and
mailing address: PO Box 439027, San Ysidro, CA 92143-9027, USA\label{8}
}

% ------------------------------
%           Abstract           
% ------------------------------
\abstract{}
{We report our finding of a high excitation emission line nebula associated with an Ultra Luminous X-ray source (ULX) at $z=$ 0.027, which we found in our \textit{Chandra} observation of the \textit{AKARI} North Ecliptic Pole (NEP) Deep Field.
} 
{We present a \textit{Chandra} X-ray and Gran Telescopio Canarias (GTC) optical spectral analysis of the ULX blob. We investigate the nature of the emission line nebula by using line ratio diagnostic diagrams, and its physical properties estimated from Spectral Energy Distribution (SED) fitting. 
}
{The optical spectrum of this ULX blob shows emission line ratios that are located on the borderlines between star-formation and Seyfert regimes in [OIII]/H$\beta$-[OI]/H$\alpha$, [OIII]/H$\beta$-[SII]/H$\alpha$ and [OIII]/H$\beta$-[OIII]/[OII] diagnostic diagrams. These are in contrast with those of a nearby blob observed with the same slit, which occupy the HII regimes. 
}
{This result suggests that the ionization of the emission line nebula associated with the ULX is significantly contributed by energy input from the accretion power of the ULX, in addition to the star formation activity in the blob, suggesting the existence of an accretion disk in the ULX emitting UV radiation, or exerting shock waves.
}

\keywords{X-rays: galaxies, individual(ANEPD-CXO104) $-$ galaxies: individual(NEP J175536.7+664546.8)}

\titlerunning{HEELN in an ULX at $z=$0.027}

\maketitle
% ------------------------------
%        Introduction
% ------------------------------
\section{Introduction}

Ultra Luminous X-ray sources (ULX) are off nucleus X-ray point sources found in galaxies that have X-ray luminosities of roughly $\approx$ 10$^{39}-$10$^{41}$ erg s$^{-1}$, which are above the Eddington luminosity of a typical stellar mass black hole (\citeauthor{Feng} for a recent review) and below those of a typical active galactic nucleus (AGN).

The most popular models of ULX involve either intermediate-mass black holes (IMBH) of $\approx$ 10$^{2}-$10$^{4}$ $M_{\odot}$ with standard accretion disks (\citeauthor{Colbert,Makishima}) or stellar mass black holes ($\sim 10 M_{\odot}$) accreting at super Eddington rates with isotropic emissions (\citeauthor{Begelman}) and/or beamed emission within a small solid angle (e.g. \citeauthor{King}). These systems, would be frequently connected with massive star-forming regions such OB associations or superstar clusters at the high-mass limit (\citeauthor{Goad,Zezas,Liu}). 

ULXs are often located inside nebulae. Optical spectral information of ULXs and the surrounding nebula is still scarce and mostly limited to nearby ($\lesssim$ 10 Mpc) ones (\citeauthor{Pakull2};\citeauthor{Abolmasov,Gutierrez,Fabrika}). A common notable feature is the HeII $\lambda$4686 emission line, along with other common emission-lines such as [OI] $\lambda$6300, indicating a high excitation UV radiation nebula and/or shocks (e.g. \citeauthor{Kaaret}). Integral field spectroscopy around the ULX Holmberg II X-1 shows that, while the HeII emission is concentrated at the ULX position, there is an extended nebulosity of emission lines, including Balmer, [OIII] and [SII] lines extending towards one side of the ULX (\citeauthor{Lehmann,Fabrika}), probably tracing an ionization cone. From emission line diagnostics of nearby ULX nebulas, Abolmasov et al. \citeyearpar{Abolmasov} argued that some of them are photoionized, while others are shock-ionized. In addition, Feng $\&$ Soria \citeyearpar{Feng} summarize information about X-ray photo-ionized and shock-ionized nebulae.

As part of our \textit{Chandra} survey in the \textit{AKARI} NEP Deep field (R.A.= 17:55:24, dec.= +66:37:23, \citeauthor{Krumpe,Matsuhara}), we have found a new ULX in the star-forming galaxy NEP J175536.7+664546.8 at $z=$ 0.027 (\citeauthor{Shim}; $D\approx$ 110 $h_{70}^{-1}$ Mpc). In this article, we present X-ray and optical spectral analysis of the ULX and an associated nebula/star cluster. In section \ref{xray} we describe the basic X-ray properties of the ULX and show results of our X-ray spectral analysis; in section \ref{optical} the optical spectrum is presented; while in section \ref{prop} we analyze the nature of the emission line nebula and star cluster associated with the ULX and discuss its physical properties. Finally, section \ref{fin} discusses the implications of our results and section \ref{sum} summarizes our conclusions. Throughout this article we have assumed a flat cosmology with $\Omega_{\rm m}=$ 0.3, $\Omega_{\Lambda}=$ 0.7 and $H_{0}=$ 70 h$_{70}$ km s$^{-1}$ Mpc$^{-1}$.

% -----------------------------
%          X-ray data
% -----------------------------
\section{X-ray properties}\label{xray}

\begin{figure}[ht]
\begin{center}
\includegraphics[width=0.45\textwidth]{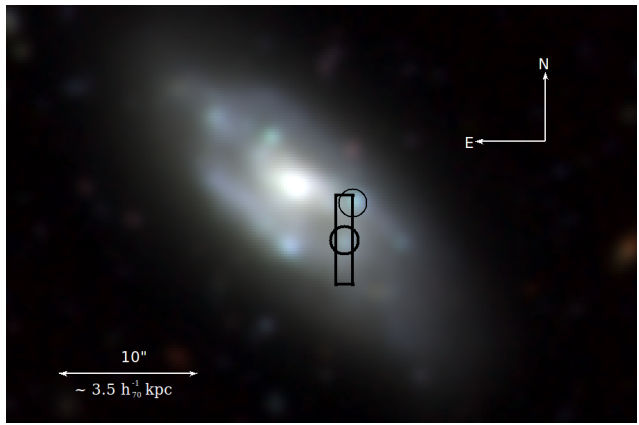}
\caption{Composite image from BRz Subaru Suprime Cam images of the galaxy NEP J175536.7+664546.8. The thick and thin circles show the positions of the ULX and nearby blobs, respectively, while the size of the circles corresponds to the aperture photometry used for the SED modelling. The rectangle approximately shows the slit geometry utilized in the OSIRIS/GTC spectroscopy.}\label{image}
\end{center}
\end{figure}

The Chandra X-ray source ANEPD-CXO104 (CID= 104 in \citeauthor{Krumpe}) is located at R.A.= 17:55:36.28, dec.= 66:45:43.8 (J2000) with a 1$\sigma$ positional error of $\approx$ 0.25$\arcsec$ (including both statistical and systematic errors). The X-ray source is co-spatial with one of the several seemingly HII-region knots along spiral structures of NEP J175536.7+664546.8, and is 5.6$\arcsec$ (projected distance of $\sim$3.1 kpc) away from the center of the galaxy. Figure \ref{image} shows an optical image of the galaxy with the position of the ULX marked with a thick circle. 

% --------- figura ULX image --------------

\begin{figure}[ht]
\begin{center}
\includegraphics[width=0.28\textwidth,angle=-90]{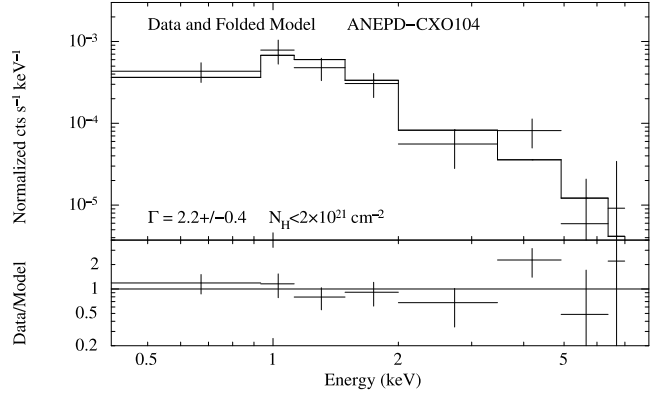}
\caption{X-ray spectrum of ANEPD-CXO104 folded with the instrumental response. The upper section shows the folded spectrum with error bars and with our best fit model. The residuals of the fit are shown in terms of data/model ratio (lower section). The error bars on data points correspond to 1$\sigma$ errors, while the error shown for $\Gamma$ is for 90$\%$ conficende level.}\label{Xspec}
\end{center}
\end{figure}

The X-ray source has $\sim$80 net counts and we only can make a minimal spectral analysis. For each OBSID, the ULX spectra are extracted from a circle centered at the source position. The size of the circle depends on the off-axis angle, source brightness and background level. To minimize fluctuations of the background level, background spectra are extracted from a large irregular shaped polygonal area on the same chip as the ULX. Care has been made to make sure no bright sources and visible structures are included in the background area. The mean off-axis angle of the area is close to that of the ULX, and the background surface brightness of various subsections are consistent with that of the background inmediately surrounding the source. The source and background extraction areas are listed in Table \ref{xobs}. Then, the spectral analysis has been made by a joint fit to four X-ray spectra of ANEPD-CXO104, accumulated from four \textit{Chandra} observations, respectively, with a simple absorbed power-law model of the photon spectrum: $P(E)\propto$ \textbf{phabs}($N_{\rm H}^{\rm Gal})\cdot$ \textbf{zphabs}($z,N_{\rm H}^{\rm src})\cdot E^{-\Gamma}$, where \textbf{phabs}\footnote{More details can be found in Xspec package manual.} represents a photoelectric absorption in our galaxy with the column density of $N_{\rm H}^{\rm Gal}=$ 4 $\times$ 10$^{20}$ cm$^{-2}$ (fixed to the Galactic value at the position of the source, \citeauthor{Kalberla}), \textbf{zphabs}($z,N_{\rm H}^{\rm src})$ is the photoelectric absorption at the redshift of the source ($z=$ 0.027), and $\Gamma$ is the photon index of the primary continuum of the source. The free parameters $N_{\rm H}^{\rm src}$ and $\Gamma$ for the all four spectra have been tied, while the normalizations have been allowed to vary individually. The fit has been made using the ``cstat'' statistic in the XSPEC\footnote{Available in https://heasarc.gsfc.nasa.gov/xanadu/xspec/} package, which fully incorporates Poisson errors of the pulse-height spectra of the source and background. As a result, the slope of the primary continuum is found to be $\Gamma=$ 2.2$\pm$0.4 (90$\%$ confidence errors). No evidence for intrinsic absorption is found, with a 90$\%$ upper limit of $N_{\rm H}<$ 2 $\times 10^{21}$ cm$^{-2}$. Also, there is no significant variability among the four observations. To obtain a representative X-ray flux, we then joined the normalizations and re-fit. The best-fit $\Gamma$ and the upper limit of $N_{\rm H}^{\rm src}$ changed very little by joining the normalization. We find  $S_{\rm 0.5-7 keV}=$ (1.4$\pm$0.3) $\times$ 10$^{-14}$ erg s$^{-1}$ cm$^{-2}$ (0.5$-$7 keV, before Galactic absorption), implying an intrinsic X-ray luminosity of $L_{\rm 0.5-7 keV}\approx$ 2.4$\pm$0.5 $\times$ 10$^{40}$ $h_{\rm 70}^{-2}$ erg s$^{-1}$ (90$\%$ confidence errors). Thus, this source can be classified as a ULX. Figure \ref{Xspec} shows the X-ray spectrum folded with the instrumental response, our best fit model and fit residuals. The summed spectrum from all four OBSIDs (rather than separate four spectra) has been rebinned to have at least 3$\sigma$ signal per bin of the source spectrum before the background subtraction, for display purposes only. The plotted model uses the best fit parameters of the joint fit with the joined normalization.

%  -------- table with X-ray observations ---------
\begin{table*}[]
\centering
\caption{X-ray observations (1$\sigma$ errors).}\label{xobs}
\begin{tabular}{c c c c c c c}
\hline\hline
OBSID & Date obs. & Exp. time & CR (0.5$-$7 keV) & Offax. & src. area & bkg. area\\
 & [Y-M-D] & [ks] & [10$^{-4}$ cts s$^{-1}$] & [arcmin] & [arcsec$^{2}$] & [10$^{4}$ arcsec$^{2}$]\\
\hline
12929 & 2011-03-16 & 11.9 & 7.2$\pm$2.5 & 2.6 & 28.3 & 2.2\\
12930 & 2011-04-01 & 14.6 & 9.7$\pm$2.7 & 6.7 & 79.2 & 7.5\\
12933 & 2010-12-28 & 23.8 & 10.0$\pm$2.0 & 6.5 & 200.9 & 4.3\\
13244 & 2011-04-02 & 14.9 & 9.4$\pm$2.7 & 7.3 & 199.0 & 2.2\\
\hline
\end{tabular}
\end{table*}

% --------- figura X-ray spectra --------------

% ---------------------------------
%          Optical data
% ---------------------------------
\section{Optical spectroscopy}\label{optical}

% --------- figura optical spectra --------------
\begin{figure*}[ht]
\begin{center}
\includegraphics[width=0.49\textwidth]{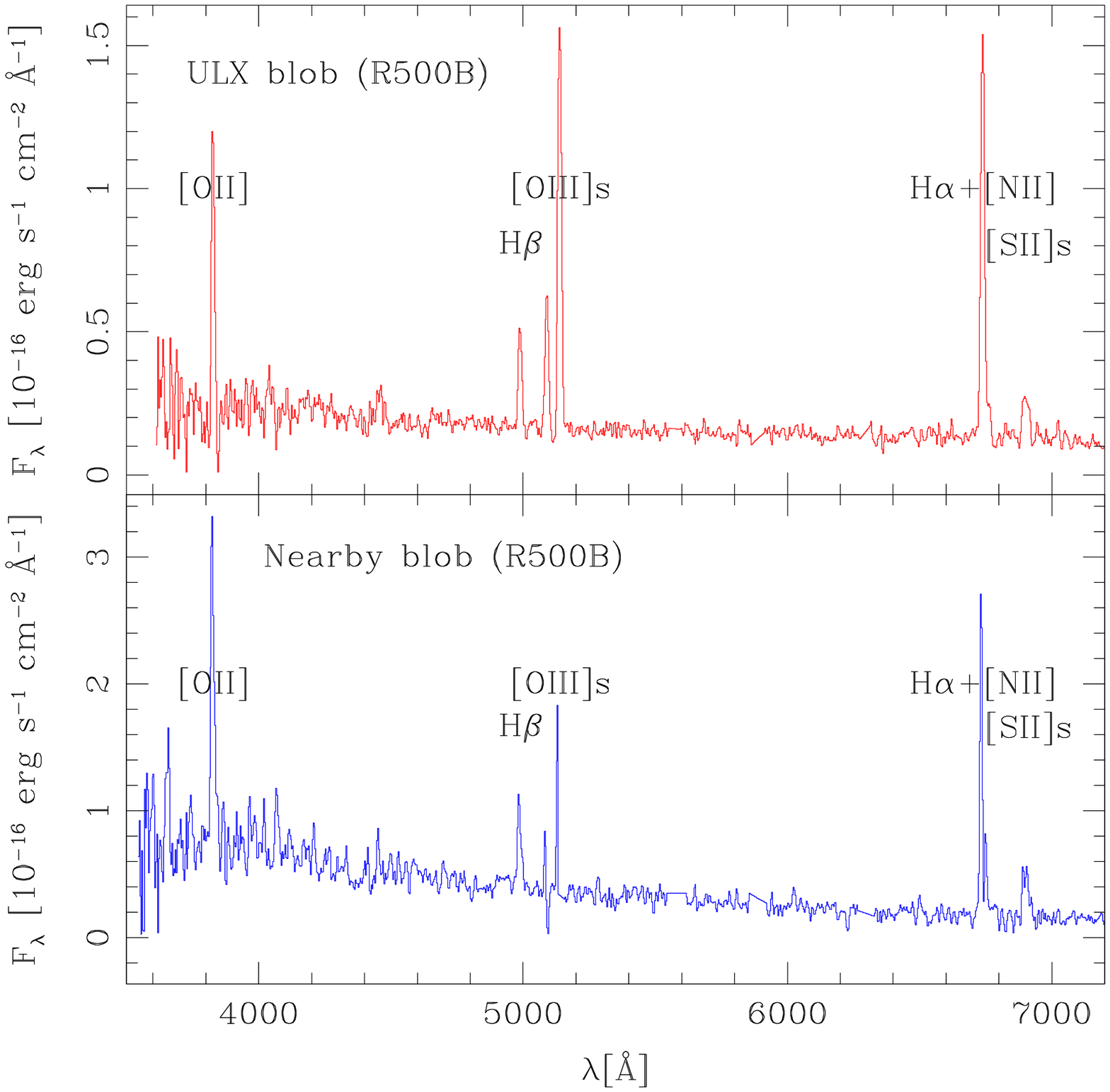}
\includegraphics[width=0.49\textwidth]{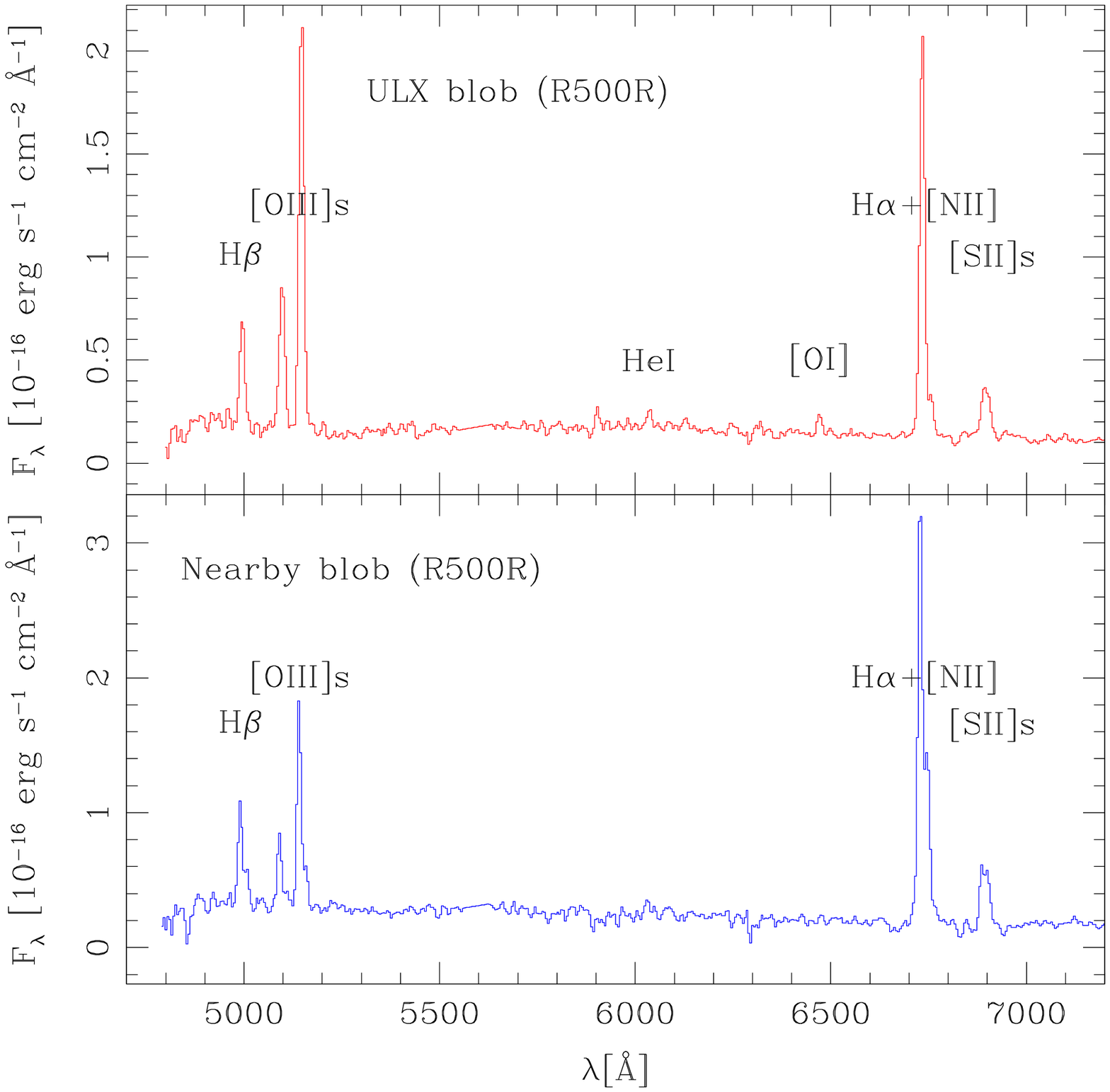}
\caption{One dimensional spectra (observed frame) of the ULX ANEPD-CXO104 (upper sections) and nearby blob (lower sections) obtained with OSIRIS/GTC. The left panel shows the spectra obtained with the R500B grism, while the right panel shows the spectra obtained with the R500R grism. Strong residual sky emission lines were removed/masked from the spectra.}\label{optspec}
\end{center}
\end{figure*}

Optical spectra were obtained for the nebula associated with our ULX ANEPD-CXO104 (hereafter referred as the ``ULX blob") with the Optical System for Imaging and low-intermediate-Resolution Integrated Spectroscopy (OSIRIS) of the Gran Telescopio Canarias (GTC), as part of a spectroscopic survey of optical counterparts of Chandra X-ray sources detected in the \textit{AKARI} NEP Deep Field. Multi-object spectroscopy was done in service mode (PI: T. Miyaji), the nights of April 23 and 26 of 2015 with the R500B and R500R grisms, which gave a dispersion of 3.54 and 4.88 $\AA/$pix, respectively. The R500B grism was utilized during the first night (Prod ID= 778892$-$778894), while the R500R grism was used during the second night (Prod ID= 779571$-$779573). With these grisms, the nominal spectral coverages are from 360 to 700 nm for R500B and from 500 to 1000 nm for R500R, respectively. However, in some cases, the R500B grism practically provided good spectra from 360 to 1000 nm, but with a lower sensitivity towards redder wavelengths. This was useful to have two measures of the features of a object. The slitwidth adopted during the observations was 1.2$\arcsec$ (geometry drawn in Fig. \ref{image}), a value higher than the average seeing of the whole run (1$\arcsec$). Integration times of 3$\times$690 s and 3$\times$1100 s were used for R500B and R500R grisms, respectively, in order to reach a signal to noise (S/N) ratio over 3 in the spectra obtained. Usual calibration frames were acquired during the run plus standard star spectra to calibrate by flux. 

The data reduction process was done using the \textbf{gtcmos}\footnote{Developed by Divakara Mayya at INAOE, Puebla, Mexico.\\Available in www.inaoep.mex/$\sim$ydm/gtcmos/gtcmos.html} package, which works in IRAF. Before using this package, the OSIRIS images composed of two CCDs with a gap and a slight shift and rotation between them, were combined into a single mosaic frame with their pixel coordinates geometrically corrected. This process was done using \textit{mosaic2x2$\_$v2}, task available on the GTC website. Then, the images were combined and bias subtracted using standard routines of IRAF. The arc line identification, wavelength calibration and sky subtraction steps were done using the \textit{omidentify, omreduce} and \textit{omskysub} routines, respectively, in order to obtain the calibrated two-dimensional spectra. The one-dimensional spectra were extracted using the \textit{omextract} routine, while the standard stars were extracted with the \textit{apall} routine. Finally, the flux calibration process was done using the \textit{standard, sensfunction} and \textit{calibrate} routines, also within IRAF.

%  -------- table with physical properties ---------
\begin{table*}[]
\centering
\caption{Physical properties}\label{tabpro}
\begin{tabular}{c c c c c c c}
\hline\hline
metallicity & object & $\tau$[Gyr] & Age[Gyr] & $E(B-V)$ & $M_{\ast}[M_{\odot}]$ & SFR[$M_\odot$/yr]\\
\hline
$Z_{\odot}$ & ULX blob & 15 & 2.00$\pm$1.22 & 0.1 & (3.63$\pm$1.13)$\times10^{7}$ & 0.022$\pm$0.008\\
& Nearby blob & $\infty$ & 1.43$\pm$0.45 & 0 & (2.39$\pm$0.30)$\times10^{7}$ & 0.021$\pm$0.003\\
\hline
0.4 $Z_{\odot}$ & ULX blob & 30 & 0.09$\pm$0.07 & 0.4 & (1.29$\pm$0.51)$\times10^{7}$ & 0.162$\pm$0.096\\
& Nearby blob & $\infty$ & 0.18$\pm$0.15 & 0.2 & (1.20$\pm$0.54)$\times10^{7}$ & 0.078$\pm$0.055\\
\hline
\end{tabular}
\end{table*}

Along with a spectrum of the ULX blob, the spectrum of another nearby blob (hereafter referred as the ``nearby blob") was also partially taken within the same slit. These objects resulted to have the same redshift of the host galaxy ($z=$ 0.027) making us also note that both object belong indeed to the galaxy. Figure \ref{optspec} shows the spectra of both objects obtained with the R500B (left panel) and R500R (right panel) gratings. We can see both sources are active regions, with strong Balmer, [OII] $\lambda$3729 and [OIII] $\lambda$4959,5007 emission lines. A quick look indicates that our ULX ANEPD-CXO104 has a Seyfert 2 like spectrum, while the nearby blob has a star-forming region like spectrum. It can also be seen the HeI $\lambda$5876 line and the [OI] $\lambda$6300 line in emission in the R500R spectrum of the ULX blob. The HeII $\lambda$4686 emission line, which is commonly observed in optical spectra of/around ULX (\citeauthor{Kuntz}) has not been detected in our spectra.

Based on the FWHMs of narrow night sky lines, the spectral resolution achieved is roughly $\lambda/\delta \lambda_{\rm FWHM}=$ 463 and 440 for the R500B (7000 \AA) and R500R (7000 \AA) respectively. For the nearby blob, a somewhat higher spectral resolution has been achieved, because the nearby blob only overlaps with the slit partially.

% --------------------------------
%             Analysis
% ---------------------------------
\section{Diagnostic of the blobs}\label{prop}

In this section, we study the properties of the ULX blob by analyzing optical-near infrared continuum and optical emission lines. These are compared with those of the nearby star-forming blob for reference.

\begin{figure}[ht]
\begin{center}
\includegraphics[width=0.5\textwidth]{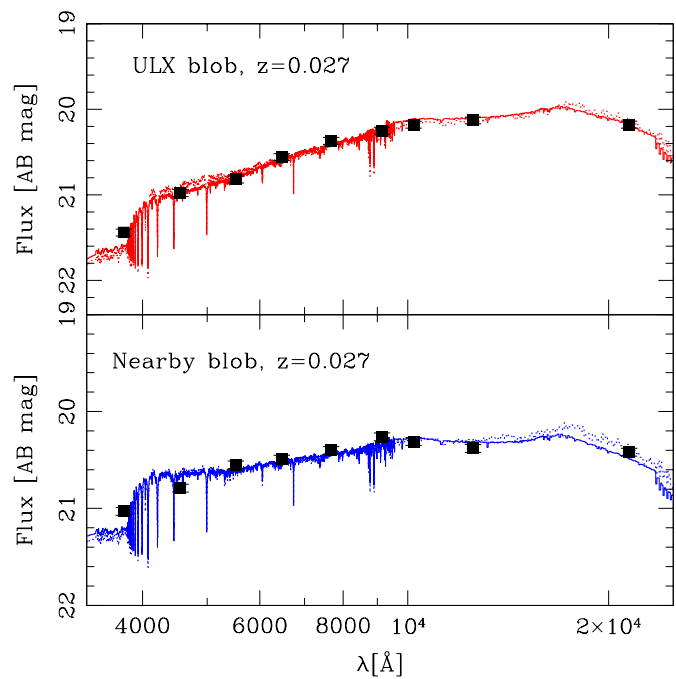}
\caption{The spectral energy distribution (SED) of the ULX (upper section) and nearby (lower section) blobs from broad band photometric data (black squares) and their best fit population synthesis models (thin lines). In both panels, the dotted line shows the fit with solar metallicity value $Z_{\odot}$, while the solid line shows the fit with metallicity 0.4 $Z_{\odot}$. The contribution from major emission lines has been subtracted from the photometric data. Error bars are about of the size of the symbols.}\label{seds}
\end{center}
\end{figure}

From the image (Fig. \ref{image}), it appears that both the ULX and nearby blobs are associated with a star-forming region/star cluster. In order to investigate the environment in which the ULX has been formed, we first studied the stellar populations of these blobs. The photometric data were extracted from broad band images obtained with CFHT Mega Cam ($u^{*}$), Subaru Suprime Cam ($B,V,R_{c},i^\prime z^\prime$), and  CFHT WIRCAM ($Y,J,K_{s}$) cameras on the \textit{AKARI} NEP Deep Field (\citeauthor{Takagi,Hanami,Murata,Oi}). A 2$\arcsec$ aperture was used for the photometry, which is shown as a circle in Fig. \ref{image}. The magnitude error associated to the aperture photometry was $\sim$0.02 mag. We subtracted the contributions from emission-lines detected in our GTC spectra from the photometric data, assuming that the rest of broad line fluxes are dominated by continuum emission from stars. In our particular case, only the photometry from $u^{*},V$ and $R$ filters needed a correction, with $\sim$0.4 mag the largest correction for $R$ filter and $\sim$0.2 mag the smallest correction for $u^{*}$. We then applied a Spectral Energy Distribution (SED) fittings using stellar population synthesis models. SED templates were generated with the GALAXEV code (\citeauthor{Bruzual}), with ages of [0, 13] Gyr and two values of metallicity, $Z_{\odot}$ and 0.4 $Z_{\odot}$ respectively. We assumed a Salpeter initial mass function (IMF,  \citeauthor{Salpeter}) and exponentially declining star formation histories (SFR(t)$\sim \exp^{-t/\tau}$), with $\tau$ values of [0,$\infty$] Gyr. A reddening law (\citeauthor{Allen}) was applied to each template with $E(B-V)$ values of [0,1]. Then, we used the photometric redshift code Hyperz (\citeauthor{Bolzonella}) for the SED fitting. Table \ref{tabpro} shows the properties estimated for each object, while Fig. \ref{seds} shows the emission-line subtracted photometries and the best fit models for each blob. The results of the SED fits show similar star formation histories between the ULX and nearby blobs at both assumed metallicities. They both shows almost constant star formation rates (SFR) with comparable ages, which are consistent with each other within errors in either choice of the metallicity. If we assume a sub-solar metallicity, the SED fits imply higher current SFR, which started more recently than the solar metallicity case. In any case, the similarity suggests that both star-forming blobs were born simultaneusly some 0.1$-$2 Gyr ago, depending on the assumed metallicity. Nevertheless, the subsolar metallicity case is likely to be correct as it is discussed in Section \ref{fin}.

% --------- figure SEDs diagram  --------------

We then analyzed the emission-line properties of the blobs observed by GTC. The flux scales of the GTC spectra have been normalized to  2$\arcsec$ aperture fluxes using the $B-, V-, R_{c}-, i-$ photometric measurements from the images described above. This procedure compensates for flux losses caused by the small slit width of the spectroscopic observations. Then the contributions from stellar population continua, which contain features like Balmer absorption, have been subtracted from the spectra using the best-fit model described above. This correction increased the measured line fluxes at most in 20$\%$ for the H$\beta$ line. Finally, emission line fluxes were measured by Gaussian fits using the \textit{splot} routine of IRAF as well as our own software. Table \ref{lines} shows the line luminosities computed for each emission line present in the spectra. The 90$\%$ confidence statistical errors or 95$\%$ upper limits are also indicated. Due to our limited resolution, H$\alpha$ and [NII$]\lambda\lambda$6548,6583 are blended. In either case, the flux of the single Gaussian fit is dominated by the H$\alpha$ side and leaves some excess residuals longward of H$\alpha$ around [NII]. Forcing a double Gaussian fit gives a small flux for [NII]. A rough estimation of the [NII]/H$\alpha$ ratio gives a value of $\sim$ 0.1. However, the line profiles are not necessarily exact Gaussian and a slight deviation from Gaussian may cause a large flux missestimation of the subdominant [NII] line. Therefore, the results of a single Gaussian fit to the H$\alpha$+[NII] feature are shown in Table \ref{lines} and we use them as representative values for H$\alpha$. Thus the H$\alpha$ luminosities shown in Table \ref{lines} are subject to contaminations from [NII]. Instead, for the R500B spectrum of the nearby blob, we were able to obtain a reasonable multi-component fit with a model including H$\alpha$ and [NII] $\lambda \lambda$6548,6583, where the flux ratio between [NII]$\lambda$6548 and $\lambda$6583 lines is fixed to the theoretical value of 3.0. For this particular case, we obtain $L(\rm{H}\alpha$)=(4.46$\pm$0.05)$\times$10$^{39}$ and $L(\rm{[NII]}\lambda$6583)=(1.04$\pm$0.05)$\times$10$^{38}$ h$_{70}^{-2}$ erg s$^{-1}$. We expect that systematic errors such as a slight differences of seeing, slit positions and background sky subtraction are larger and might have caused the differences between the same lines observed with R500B and R500R gratings.
% For reference, we have included the line luminosities for the host galaxy, derived from Shim et al. \citeyearpar{Shim} data.

% --------- figure BPT diagram  --------------
\begin{figure*}[]
\begin{center}
\includegraphics[width=0.32\textwidth]{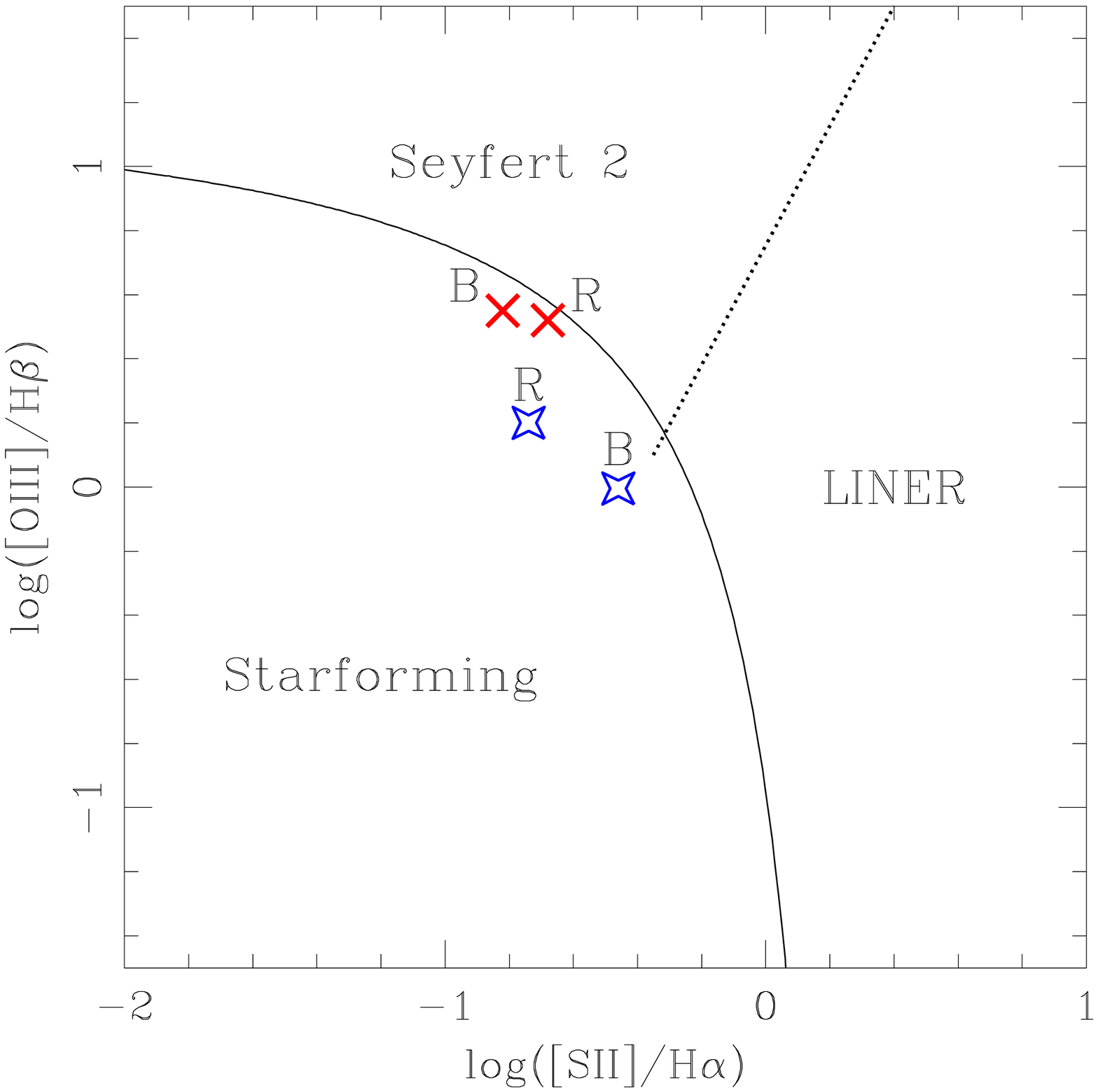}
\includegraphics[width=0.32\textwidth]{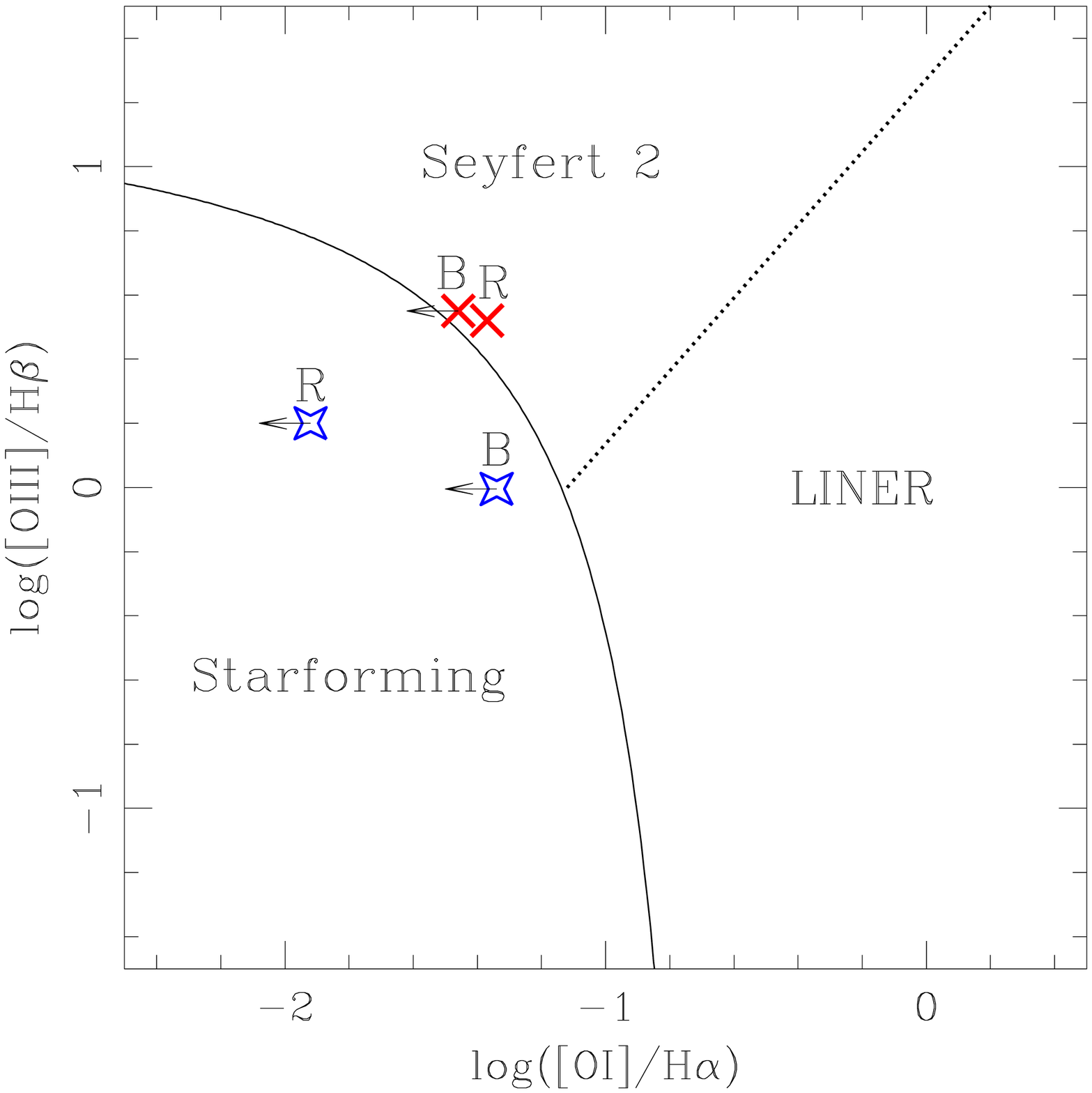}
\includegraphics[width=0.32\textwidth]{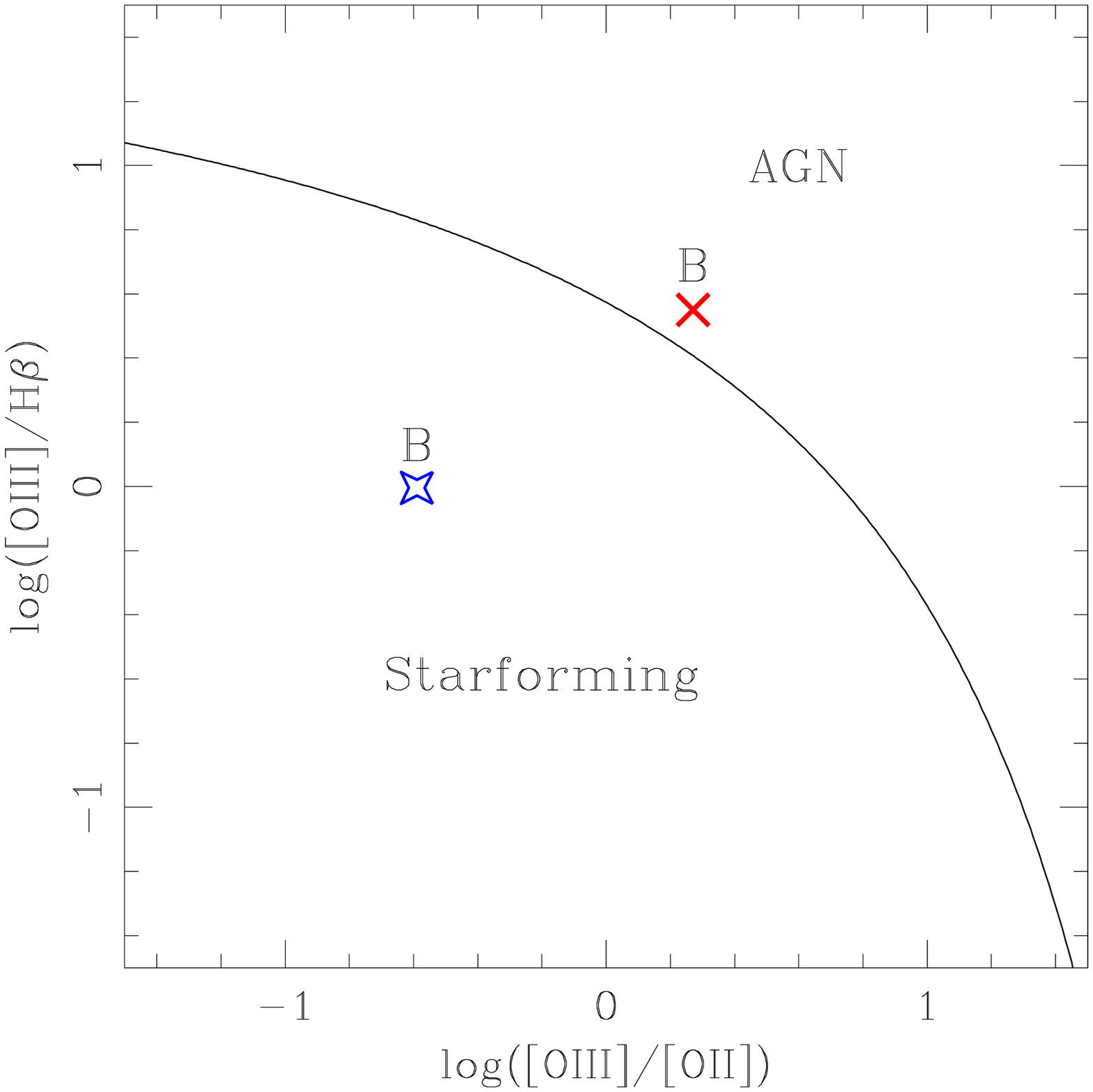}
\caption{Diagnostic log([OIII]/H$\beta$) vs log([SII]/H$\alpha$) (left), log([OIII]/H$\beta$) vs log([OI]/H$\alpha$) (middle) and log([OIII]/H$\beta$) vs log([OIII]/[OII]) (right) diagrams. In the left and middle diagrams, the boundary lines were taken from Kewley et al. \citeyearpar{Kewley}, while in the right diagram the boundary line was taken from Evans $\&$ Dopita, \citeyearpar{Evans}. The ULX blob is represented with a red cross, while the nearby blob is shown with a blue star. The B symbol indicates the line ratios computed from R500B grism, while the R shows the line ratios computed from R500R grism. Statistical errors are smaller than the symbol sizes, except those for log([OI]/H$\alpha$).}\label{bpt} %As reference, the host galaxy is shown with an empty triangle.}\label{bpt}
\end{center}
\end{figure*}

%  -------- table with line luminosities ---------
\begin{table*}[]
\centering
\caption{Emission line luminosities [10$^{39}$ h$^{-2}_{70}$ erg s$^{-1}$] (90$\%$ statistical errors or 95$\%$ upper limits).}\label{lines}
\begin{tabular}{c c c c c c c c c}
\hline\hline
grism & object & L([OII]) & L(H$\beta$) & L([OIII]$^{1}$) & L([OIII]$^{2}$) & L([OI]) & L(H$\alpha$+[NII]) & L([SII])\\
\hline
R500B & ULX blob & 2.18$\pm$0.03 & 1.16$\pm$0.05 & 1.34$\pm$0.02 & 4.15$\pm$0.05 & $\leq$0.16 & 4.56$\pm$0.06 & 0.70$\pm$0.07\\
 & Nearby blob & 6.34$\pm$0.32 & 2.18$\pm$0.24 & 0.55$\pm$0.11 & 1.62$\pm$0.08 & $\leq$0.21 & 4.53$\pm$0.09 & 1.59$\pm$0.12\\
\hline
R500R & ULX blob & $-$ & 1.67$\pm$0.06 & 2.03$\pm$0.06 & 5.58$\pm$0.06 & 0.22$\pm$0.04 & 5.30$\pm$0.05 & 1.13$\pm$0.04\\
 & Nearby blob & $-$ & 2.10$\pm$0.12 & 1.5$\pm$0.10 & 3.35$\pm$0.06 & $\leq$0.13 & 10.56$\pm$0.10 & 1.87$\pm$0.07\\
% host$^{5}$ & 10.87 & 1.84 & 1.47 & 3.90 & 0.38 & 11.33 & 4.35\\
\hline
\end{tabular}
\footnotesize{\\Note: 1. [OIII] $\lambda$4959, 2. [OIII] $\lambda$5007.} %5. Data derived from Shim et al. \citeyearpar{Shim}.}
\vspace{0.3cm}
\end{table*}

In order to see if the emission-lines of the ULX nebula are solely ionized by OB stars or there is significant contribution from another source, we plot line ratios on two diagnostic diagrams that are customarily used to classify narrow emission line galaxies into Seyferts, LINERs, and HII galaxies. We are unable to use the most common diagram with log([OIII]/H$\beta$) vs log([NII]/H$\alpha$) due to the blending problem. However, Figure \ref{bpt} (left, middle and right panels) shows the Kewley et al. \citeyearpar{Kewley} log([OIII]/H$\beta$) vs log([SII]/H$\alpha$) (left), log([OIII]/H$\beta$) vs log([OI]/H$\alpha$) (middle) and the Evans $\&$ Dopita \citeyearpar{Evans} log([OIII]/H$\beta$) vs log([OIII]/[OII]) (right) star-forming-AGN diagnostic diagrams for our ULX blob. It can be seen that the ULX blob (red crosses) is located on the borderlines between star-formation and Seyfert/AGN regimes in all diagrams. Only a few percent of HII/star-forming galaxies without AGNs from Veilleux $\&$ Osterbrock \citeyearpar{Veilleux} data fall into the upper-right quadrant with respect to the ULX blob position. On the other hand, the nearby blob (blue star) shows a line ratio typical of a star-forming regions in both diagnostic diagrams. We have estimated an upper limit value for the [OI] luminosity of the nearby blob, according to the limit flux measurable in their spectra. The [OI] upper limits for the nearby blob are also plotted in Fig. \ref{bpt} and listed in Table \ref{lines}. Thus, the presented results suggest a noticeable contribution from accretion process to the ionizing radiation, producing the emission line nebula in the ULX blob.

% -------------------------------
%         Discusion
% -------------------------------
\section{Discussion}\label{fin}

The star formation rate and stellar masses of the ULX and the nearby blobs are very similar. The star formation rates inferred by the SED fits ($\sim$ 0.02$-$0.16 M$_{\odot}$ yr$^{-1}$) are close to those estimated by their H$\alpha$ and [OII] line luminosities ($\sim$ 0.02$-$0.08 M$_{\odot}$ yr$^{-1}$), using the Kennicutt et al. \citeyearpar{Kennicutt} formulas. The ULX thus resides in an actively star-forming region. Our slit size of 1.2$\arcsec$ corresponds to $\sim$ 650 $h_{70}^{-1}$ pc ($D\approx$ 110 $h_{70}^{-1}$ Mpc), while the emission-line regions of nearby ULX nebulas typically extend to $\lesssim$ 500 pc (\citeauthor{Pakull,Abolmasov}, hereafter A07). The stellar masses derived from the SED fits, $\sim$(1.2$-$3.6)$\times$10$^{7}$ M$_{\odot}$, are much higher than typical stellar masses of OB associations (10$^{2}-$10$^{4}$ M$_{\odot}$, \citeauthor{Fall}). However, star clusters with higher stellar mass values has been inferred to exist in the Galaxy ($\sim$10$^{5}$ M$_{\odot}$, \citeauthor{Rahman}) and other external galaxies ($\sim$10$^{6}$ M$_{\odot}$, \citeauthor{Larsen,Weidner}). Then, the stellar mass estimated for the blobs might correspond with a massive star cluster composed of a group of OB associations. Thus, we would expect that the emission line spectra of the ULX blob in our observations consist of a mixture from HII regions ionized by OB stars and a nebula ionized by the ULX.

\defcitealias{Abolmasov}{A07}

In order to see if it is reasonably to expect the contribution from accretion, we compare the [OIII]$\lambda 5007$ to X-ray (0.3$-$10 keV) luminosity ratio of the ULX blob with those of known ULX nebulas. The $L_{\lambda 5007}/L_{0.3-10\rm keV}\approx$ 0.15 for our ULX blob is comparable to those of two of the ULX nebulas studied by \citetalias{Abolmasov}, namely NGC6946 X-1 and NGC7331 X-1, with $L_{\lambda 5007}/L_{0.3-10\rm keV}\approx$ 0.1 and 0.06 respectively (0.3$-$10 keV luminosities are from \citeauthor{Swartz}). Furthermore, even with a low value of [NII]/H$\alpha \sim$ 0.1 for our ULX blob, which has been obtained with an unreliable fit in Section \ref{prop}, the ULX contribution is still pausible. In fact, the ULX nebula associated with NGC5204 X-1 by \citetalias{Abolmasov} has a high value of [OIII]/H$\beta \sim$ 5 and still has a rather low value of [NII]/H$\alpha \sim$ 0.13. Low-mass/low-metallicity photoionized or shock-heated nebulae are expected to show low values of the [NII]/H$\alpha$ ratio, while keeping the [OIII]/H$\beta$ ratio relatively unchanged (\citeauthor{Groves,Dopita}). Thus, we may reasonably expect significant contribution of the ULX accretion to the line ratios, which pushes the ULX nebula towards the Seyfert regime in the emission line diagnostic diagrams. In addition, as it was mentioned above, the low stellar mass estimated for our ULX nebula and the low [NII]/H$\alpha$ ratio make us to prefer the subsolar metallicity value used in the SED fits with its derived properties as the most likely (Table \ref{tabpro}).

It is not surprising that we did not detect HeII $\lambda$4686 emission, which is common in nearby ULXs. Our line detection upper limit is $L_{\rm lim}\sim$ 3$\times$10$^{38}$ h$_{70}^{-2}$ erg s$^{-2}$, while the range of the HeII line luminosities for the nearby nebulae is $L_{\rm HeII}\sim$ (1$-$11)$\times$10$^{36}$ h$_{70}^{-2}$ erg s$^{-2}$ \citepalias{Abolmasov}.

While our observational clues to the nature of the ULX, especially in terms of the constrains on the mass of the black hole of the ULX, are limited, it might be instructive to present a few possible cases that are consistent with our observations and to discuss their implications. The ionization of the ULX nebula may be either photoionization or heating by a shock with velocity $V\gtrsim$ 100 km s$^{-1}$ (e.g. \citeauthor{Allen2,Dopita}). Assuming a photoionization picture, let $f$ be the fraction of H$\beta$ emission that is contributed by the ULX accretion. Naturally, the value of $f$ is essentially unconstrained. Since the SFR (and associated H$\alpha$ and H$\beta$ fluxes) from the SED fits varies by a factor of several as we vary the assumed metalicity, a reasonable estimate of $f$ can not be obtained by comparing the Balmer fluxes expected from the SFR from the population synthesis results and from the ULX blob. Nevertheless, the $L_{\rm [OIII]}/L_{\rm H\beta}$ values of nearby ULX nebulae range from $\sim$ 2$-$7 (\citetalias{Abolmasov}). For example, $L_{\rm [OIII]}/L_{\rm H\beta}\sim$ 7 for the case with high ULX ionization (from NGC6946 X-1 observation) and $L_{\rm [OIII]}/L_{\rm H\beta} \sim$ 2 (from our nearby nebula) for the case with OB star ionization. Then, we obtain $f \sim$ 0.35 to have $L_{\rm [OIII]}/L_{\rm H\beta} \sim$ 3.5, which is the observed value for our ULX blob. Instead, if we assume $L_{\rm [OIII]}/L_{\rm H\beta} \sim$ 4 for our ULX nebula, $f \sim$ 0.75.  Here we discuss the implications in the following cases of accretion.

\begin{enumerate}
\item [a)] \textit{The ULX accretion is Eddington-limited (e.g. case of a standard $\alpha$-disk, \citeauthor{Shakura}) and the ULX nebula is shock-heated.} In this case, we can assume the bolometric luminosity is dominated by the X-ray emission. Then, $L_{\rm bol}\sim$ 2$\times$10$^{40}$ erg s$^{-1}$, and the black hole mass (BH) would be $M_{\rm BH}>$ 140 M$_{\odot}$.

\item [b)] \textit{The ULX accretion is Eddington-limited and the ULX nebula is photo-ionized.} Under this picture, the eq. (5) of \citetalias{Abolmasov} implies an ionizing UV radiation from ULX accretion disk with $L_{\rm UV}\sim$ 1.2$\times$10$^{41}f$ h$_{70}^{-2}$ erg s$^{-1}$. We also have to consider the ionizing UV luminosity in addition to the X-ray luminosity, and therefore the bolometric luminosity would become $L_{\rm bol}\sim$ $(2+12f)\times$10$^{40}$ erg s$^{-1}$. The corresponding BH mass would be $M_{\rm BH}>$ 140$(1+6f)$ $M_{\odot}$ in order not to violate the Eddington-limit. If $f=0.5$, the BH mass would be $M_{\rm BH}\gtrsim$ 1100 M$_{\odot}$.

\item [c)] \textit{The ULX accretion is a super critical accretion disk (SCAD) and the ULX nebula is shock-heated.} The SCAD is either one of those in which most of the accreted energy is advected onto the black hole (slim disk, \citeauthor{Watarai}) or those in which a bulk of energy output goes to outflows (\citeauthor{Poutanen}). Under SCAD models a radiation output (true luminosity) can exceed the Eddington limit by a factor 2$-$3. A further boosting of a factor of a few is possible for an apparent luminosity if it is combined with a mild collimation  (\citeauthor{Feng} and references therein). Thus, in the SCAD model with a shock heating of the ULX nebula, the apparent luminosity can be $\sim$ 10 times higher than the corresponding Eddington limit. Since the observed X-ray luminosity is expected to be the bolometric luminosity, the BH mass can be as low as $M_{\rm BH}\gtrsim$ 15 M$_{\odot}$, which is consistent with a stellar mass BH.

\item [d)] \textit{The ULX accretion is a SCAD and the ULX nebula is photo-ionized.} In this case, we should treat the X-ray and UV luminosities differently. The mild collimation boost only applies to the apparent X-ray luminosity, but does not apply to the UV luminosity, which is implied by the H$\beta$ emission. If the SCAD allows to emit $b_{\rm S}$ times more radiation than the Eddington luminosity and the collimation boosting factor is $b_{\rm C}$, the implied BH mass becomes $M_{\rm BH}\gtrsim$ 140 $(1/b_{\rm C}+12f)/b_{\rm S}$ M$_{\odot}$. If we assume $b_{\rm S}\approx$ $b_{\textbf C}$ $\approx$ 3 and $f=$ 0.5, then $M_{\rm BH}\gtrsim$ 300 M$_{\odot}$.
\end{enumerate}

Under the scenarios (b) and (d), the existence of an IMBH ($M_{\rm BH}\sim$ 10$^2-$10$^4$ M$_{\odot}$) is implied. A major problem about the IMBH picture is that there is no known black hole in this mass range, while it has been theoretically suggested that the IMBHs can form from collapsed massive population III stars or at the nucleus of star clusters (\citeauthor{Feng} and references therein). On the other hand, for $f\sim$ 0.1 under the scenario (d), the inferred BH mass is $M_{\rm BH}\gtrsim$ 70 M$_{\odot}$. There is at least one known example of a black hole with a similar mass, which is the remnant of the black hole merger gravitational wave source GW150914 ($M_{\rm BH}=$ 62$\pm$4 M$_{\odot}$, \citeauthor{Abbott}). Finally, the existence of a couple of ULXs powered by an accreting neutron star should also be mentioned. In this scenario, a neutron star reaches a $L_{X}>$ 10$^{40}$ $h^{-2}_{70}$ erg s$^{-1}$ accreting at super-Eddington rates in a binary system (M82 X2, \citeauthor{Bachetti}; NGC 7793 P13, \citeauthor{Furst,Israel}).

% ------------------------
%         Summary
% ------------------------
\section{Conclusions}\label{sum}

We have discovered a new ULX at $z=$ 0.027, with \textit{Chandra} observations in the AKARI NEP Deep field. The ULX, ANEPD-CXO104 is located $\sim$ 3.1 kpc away from the center of the galaxy NEP J175536.7+664546.8, and it is associated to a high excitation emission line nebula. The location of the ULX coincides with one of several HII region blobs along spirals of the galaxy.

The X-ray spectrum shows a power-law shape with slope $\Gamma=$ 2.2$\pm$0.4 without evidence for intrinsic absorption.  The X-ray luminosity estimated for our ULX is $L_{\rm 0.5-7 keV}\approx$ 2.4$\pm$0.5$\times$10$^{40}$ $h_{70}^{-2}$ erg s$^{-1}$.

The optical spectrum of the ULX blob shows emission lines typical of active regions with strong Balmer, [OII] and [OIII] emission lines. The [OI] $\lambda$6300 and HeI $\lambda$6876 lines are also detected. The spectrum of other HII blob nearby was also taken within the same slit, presenting the same redshift and similar spectral features.

We applied a SED fitting to the photometric data. The fits for both blobs revealed very similar properties, well represented by a continuous, almost constant star formation rate (SFR, $\sim$ 0.02$-$0.16 M$_{\odot}$ yr$^{-1}$), which started 0.09$-$2 Gyr ago. These SFR values are close to those estimated by their H$\alpha$ and [OII] luminosities ($\sim$ 0.02$-$0.08 M$_{\odot}$ yr$^{-1}$). The stellar masses estimated for both blobs ($\sim$1.2$-$3.6$\times$10$^{7}$ M$_{\odot}$) are much higher than typical values of OB associations (10$^{2}-$10$^{4}$ M$_{\odot}$, \citeauthor{Fall}), and they might correspond with massive star clusters composed of a group of OB associations. 

In order to see if the emission lines of the ULX nebula are produced solely by OB stars or by contribution from other source, we used three diagnostic diagrams: the log([OIII]/H$\beta$) vs log([SII]/H$\alpha$) and log([OIII]/H$\beta$) vs log([OI]/H$\beta$) diagrams (from \citeauthor{Kewley}), and the log([OIII]/H$\beta$) vs log([OIII]/[OII]) diagram (from \citeauthor{Evans}). The ULX blob is located on the borderlines between star formation and Seyfert/AGN regimes in all diagrams. These results are in contrast with those of the nearby blob, which occupy the HII regimes.

The ratio between [OIII] $\lambda$5007 and X-ray 0.3$-$10 keV luminosities is $L_{\lambda 5007}/L_{\rm 0.3-10  keV}\approx$ 0.15 for our ULX blob, a comparable value to those of other ULX nebulae (\citeauthor{Abolmasov}). Thus, based on our results we may reasonably expect significant contribution of the ULX accretion to the emission lines, moving the location of the ULX nebula towards the Seyfert regime in the diagnostic diagrams.

Due to the distance to our ULX, the analysis are limited by the spatial resolution. High spatial resolution spectroscopy such as available with the Space Telescope Imaging Spectrograph (STIS) attached to the Hubble Space Telescope (HST) would give a superb emission-line diagnostics of this ULX, with much less contamination from the surrounding star-forming regions.

% ----------------------------------------
%         Acknowledgements
% ----------------------------------------
\begin{acknowledgements}

This research was partially supported by Consejo Nacional de Ciencia y Tecnolofía (CONACyT, México; grants 179662 and 252531) and the Dirección General de Asuntos del Personal Académico (DGAPA-UNAM, México) through the PAPIIT program (IN104216). JD thanks to DGAPA for the fellowships awarded to do this research. TI, HH and YU thank financial support by the Grant-in-Aid for Scientific Research (26400216, 21340042, 24650145, 26400228) from Japan Society for the Promotion of Science (JSPS). MK acknowledges financial support by DFG grant KR3338/3-1.

This study is based on data principally collected using the Gran Telescopio de Canarias, operated by the Instituto de Astrofísica de Canarias, under active support of the Spanish Government and the Local Government from the Canary Islands through the European Funds for the Regional Development (FEDER) provided by the European Union, including participation of the Instituto de Astronomía de la Universidad Nacional Autónoma de México (IA-UNAM), the Instituto Nacional de Astrofísica, Óptica y Electrónica (INAOE, México) and the University of Florida (United States); the Chandra X-ray Observatory, operated by the Smithsonian Astrophysical Observatory for and on behalf of the National Aeronautic Space Administration (NASA); the Subaru telescope operated by the National Astronomical Observatory of Japan (NAOJ) and the Canada-France-Hawaii Telescope (CFHT) operated by the National Research Council (NRC) of Canada, the Institut National des Sciences de l'Univers of the Centre National de la Recherche Scientifique of France, and the University of Hawaii.
 
%and the AKARI space mission for infrared astronomy, operated by the Japan Aerospace Exploration Agency (JAXA) with the participation of European Space Agency (ESA);
\end{acknowledgements}

% -----------------------------
%       Bibliography
% -----------------------------

\end{document}